\documentclass[twocolumn,showpacs,prl]{revtex4}

\ifx\pdfoutput\undefined
\usepackage{graphicx}
\else
\usepackage[pdftex]{graphicx}
\usepackage{epstopdf}
\fi

\usepackage[center]{subfigure}
\usepackage{bm}
\usepackage{latexsym}

\def\ie{{\it i.e.,\ }}

\def\beq{{\begin{equation}}}
\def\eeq{\end{equation}}

\begin{document}

\title{Critical behavior of Griffiths ferromagnets}
\author{Pak Yuen Chan, Nigel Goldenfeld and Myron Salamon}
\affiliation{Department of Physics, University of Illinois at
Urbana-Champaign, Loomis Laboratory of Physics, 1110 W. Green St.,
Urbana, IL 61801-3080, USA}

\date{\today}

\begin{abstract}
From a heuristic calculation of the leading order essential singularity
in the distribution of Yang-Lee zeroes, we obtain new scaling relations
near the ferromagnetic-Griffiths transition, including the prediction
of a discontinuity on the analogue of the critical isotherm.  We show
that experimental data for the magnetization and heat capacity of
$\mathrm{La_{0.7}Ca_{0.3}MnO_3}$ are consistent with these predictions,
thus supporting its identification as a Griffiths ferromagnet.

\end{abstract}

\pacs{75.40.Cx, 75.40.-s, 75.47.Lx}

\maketitle

The influence of disorder on ferromagnets remains, after more than 30
years of effort, a complex and poorly-understood phenomenon.  In its
simplest form, disorder can be represented as a random spatial
variation of the exchange interaction $J$ in the bonds between
neighbouring sites on a regular lattice.  If a great enough fraction $p
> p_c$ of the bonds have $J=0$, then one would expect that there is a
vanishingly small probability of finding a percolating pathway of bonds
throughout the system, and the cooperative ferromagnetic phase would
cease to exist.  For smaller values of $p$, we would expect that the
ferromagnetic phase will exist in a form weakened by the shortage of
percolating paths; hence thermal fluctuations will destroy the
ferromagnetic phase at a temperature $T_c$ which is {\it lower\/} than
the critical temperature $T_G$ of the pure ferromagnet.  However, as
Griffiths showed\cite{Griffiths} it is not the case that the phase for
$T_c < T < T_G$ is purely paramagnetic, because in the thermodynamic
limit, there can exist arbitrarily large volumes of the system that are
devoid of disorder, with a probability exponentially sensitive to the
volume.  As a result, the free energy is non-analytic in external
field, $h$, throughout the whole Griffiths phase.  The effect of
disorder is to partition the pure system into small ferromagnetic
clusters.  Depending on the size, each cluster has a different value of
$T_c$, so that the system as a whole exhibits a spectrum of $T_c$,
spanning from the critical temperature of the pure system, $T_G$, due
to arbitrarily large clusters, to some value of $T_c$, contributed by
smaller clusters.

Here we are concerned with the phase transition between the
ferromagnetic and Griffith's phases.  Just as in the case of a pure
ferromagnet, one would like to predict the critical phenomena, but the
non-analytic nature of the Griffith's phase makes it difficult to apply
off-the-shelf renormalization group techniques\cite{harris1974rga,
khmelnitskii1975pts, grinstein1976arg} or to posit simple scaling laws,
despite recent theoretical progress\cite{Bray, Bray2, Bray3,
ruizlorenzo1997gst, Dotsenko99, Dotsenko06}. Indeed, it is currently
controversial whether or not there is clear experimental
evidence\cite{Salamon02, Li, deisenhofer2005ogp, Magen06} supporting
the existence of the Griffiths phase.  From the practical perspective,
perhaps the most unsatisfactory aspect of efforts to relate theory to
experiment is that critical exponents derived from conventional scaling
laws are unrealistically large: for example, the critical isotherm
exponent was recently\cite{Salamon03} estimated as $\delta=17$.  The
breakdown of conventional scaling strongly suggests that the functional
form of the scaling relations actually reflects the essential
singularities intrinsic to the Griffiths phase, and that some form of
exponential scaling, rather than algebraic scaling, is appropriate.

The purpose of this Letter is to address these problems by exploring
the expected form of scaling relations that would follow from a leading
essential singularity contribution to the statistical mechanics in the
Griffiths phase.  Such a contribution can be conveniently represented
using the Yang-Lee theory of phase transitions\cite{Yang} to derive the
scaling behavior in the Griffiths phase from a simple,
physically-motivated ansatz for the distribution of partition function
zeros, following arguments originally due to Bray and
Huifang\cite{Bray} for the case of long-ranged ferromagnets.  We
demonstrate that the leading singularities in the thermodynamics can be
deduced, and that our predictions consistently describe high quality
magnetic and thermodynamic data\cite{Salamon02, Salamon03} on the
disordered Heisenberg ferromagnet La$_{0.7}$Ca$_{0.3}$MnO$_3$.  We
emphasize that our purpose is only to identify the leading essential
singularities, and that it is beyond the scope of our work to provide a
full description valid outside of the asymptotic critical regime.
Nevertheless, the experimental data have sufficient resolution that we
have strong support for our scaling predictions in this asymptotic
regime.

Magnetic properties are well-accounted for by our approach, but heat
capacity data are not expected to follow a simple scaling form--and
indeed do not--due to the complex form of the theoretical predictions
which arise from even the simplest Yang-Lee zero distribution function
that we use. The non-analyticity of magnetization in external field - a
signature of the Griffiths phase - is also explicitly demonstrated.  Our
results provide strong evidence for a Griffiths singularity, and
highlight the need for a more systematic renormalization group approach
to understanding the singularities in such disordered systems.

\bigskip
\noindent
{\it Yang-Lee zeroes and critical phenomena:-\/}
In 1952, Lee and Yang\cite{Yang} developed a theory of phase
transitions based upon the density $g$ of zeroes of the grand
partition function as a function of the complex fugacity and showed
that the zeroes lie on a unit circle in the complex plane,
parameterized below in terms of the angle $\theta$. The distribution
$g(\theta)$ varies with temperature $T$ and dictates the functional
form of thermodynamics. Near a critical point, it is expected that
$g(\theta, T)$ exhibits behavior which reflects the non-analyticity
of the thermodynamics\cite{Itzykson,Binek}, and we provide this
connection explicitly here, for both the case of conventional
ferromagnetic critical point scaling, and then for the scaling near
the Griffiths point.

We begin with the scaling of the magnetization per spin, $M(H,t)$,
where $H$ is the external magnetic field and $t\equiv (T-Tc)/Tc$
is the reduced temperature, for a regular Ising ferromagnet.
The exact relationship between $M(H,t)$ and $g(\theta,t)$ can be written
as\cite{Yang}
\begin{equation}\label{long_int_eqn}
M(H,t)=2 \mu \int_0^\pi d\theta \, \frac{g(\theta,t) \tanh(\mu H/k_B T)
[{1+\cot^2(\theta/2)}]}{1+(\tanh(\mu H/k_B T)\cot(\theta/2))^2},
\end{equation}
where $\mu$ is the magnetic moment of individual spins.
To extract the scaling behavior, we proceed by
expanding around the critical point $H=0$ and $t=0$.  Because the
singular behavior arises from the limit $\theta\rightarrow 0$, we
also expand the integrand about $\theta=0$, resulting in
\begin{equation}\label{int_eqn}
M(H,t)= 2\mu \frac{\mu H}{k_B T} \int_0^\pi d\theta \frac{g(\theta,t)}
{(\mu H/k_B T)^2+\theta^2/4},
\end{equation}
which is valid near the critical point, up to corrections reflecting
a smooth background.  A change of variables, $\theta=\mu H
\phi/k_BT$, gives
\begin{equation}\label{final_int_eqn}
M(H,t)=4\mu \int_0^{{\pi k_B T}/{2\mu H}} d\phi \frac{g(2\mu H\phi/k_B T,t)}{1+\phi^2}
\end{equation}
where the upper limit is replaced by $\infty$ as $H\rightarrow 0$. This
is the primary relation between $M(H,t)$ and $g(\theta,t)$ near the
critical point.  The scaling form of the magnetization,
\begin{equation}\label{M_scaling}
M(H,t)=|t|^\beta f_M(H/|t|^{\beta\delta}),
\end{equation}
where $f_M(x)$ is an unknown scaling function, and $\beta$ and $\delta$
are two critical exponents, implies a scaling form for $g(\theta, t)$.
Substituting Eqn.~(\ref{M_scaling}) into Eqn.~(\ref{final_int_eqn}), Mellin
transforming the expression and using the corresponding
convolution relation, we arrive at the
scaling form of $g(\theta,t)$ as $\theta\rightarrow 0$
\begin{equation}\label{g_general}
g(\theta,t)=|t|^\beta G(\theta/|t|^{\beta\delta}),
\end{equation}
where $G(x)$ is a scaling function for $g(\theta,t)$.

Apart from exhibiting the scaling form of $g(\theta,t)$ for normal
ferromagnets, this exercise also shows that knowledge of $g(\theta,t)$
can be gained by studying the scaling form of $M(H,t)$, and vice versa.
In the following section, we study the scaling behavior of $M(H,t)$ in
the Griffiths phase using a heuristically-derived form for $g(\theta)$.
We will see that the result is different from that of the pure case,
reflecting the intrinsic essential singularity that characterizes
Griffiths phases.

\begin{figure}[t]
\begin{center}
\includegraphics[width=0.45\textwidth]{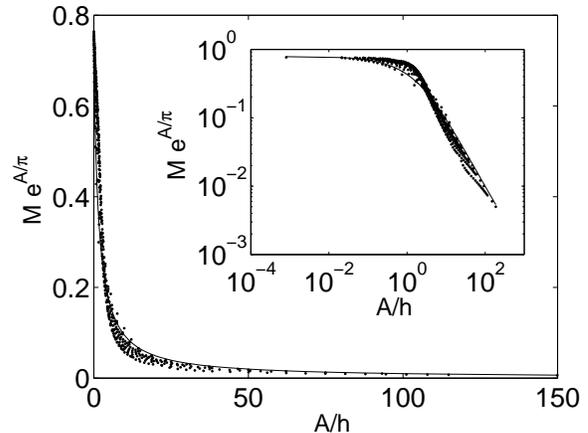}
\caption{Data collapse of magnetization, $M(h,t)$.
Dots are experimental data and the line is the
theoretical prediction for the universal scaling
function, with $g_0=0.5$, $A_0=0.5$ and
$\beta_r=0.8$. Data shown are in the range of
$t<0.35$ and $h\equiv \mu H/k_B T_c<0.39$.  The
insert shows the data on logarithmic scales.}
\label{data_collapse}
\end{center}
\end{figure}

\bigskip
\noindent {\it Density of zeros for a disordered ferromagnet:-\/} We
start with the scaling form of $g(\theta,t)$ derived on the basis of
heuristic arguments by Bray and Huifang\cite{Bray} for disordered
ferromagnets with short-ranged interactions:
\begin{equation}
g(\theta,t)=\frac{1}{\pi}\Re\sum_{r=1}^\infty \exp(-A(t)r)
\tanh\left[r(i\theta + \epsilon)\right]
\end{equation}
where $\epsilon\rightarrow 0^+$ and $A(T)\sim (T-T_c)^{2-\beta_r}$ as
$T\rightarrow T_c$ and $A(T)\rightarrow\infty$ as $T\rightarrow T_G$.
The exponent $\beta_r$ is the order parameter exponent for the random
case, and its value will reflect the universality class of the magnet,
be it Ising, Heisenberg, O($n$) etc.  In the limit that
$\theta\rightarrow 0$ and $A\rightarrow 0$, the summand is dominated by
peaks at values of $r=(2n+1)\pi/2\theta$ for all non-negative values of
$n$, whose height is given by $(2\theta/\epsilon (2n+1)\pi)\exp
(-(2n+1)\pi A/2\theta)$.  The width of these peaks therefore scales in
the same way as their separation, both being proportional to
$1/\theta$.  Thus, the expression takes the form:
\begin{equation}\label{Bray_form}
g(\theta,t)=g_0\exp(-A(t)/|\theta|),
\end{equation}
where $g_0$ is a constant. The essential
singularity in Eqn.~(\ref{Bray_form}) reflects the Griffiths phase of
disordered ferromagnets, not present in the pure case.  Disordered
magnets with long-range interactions have a power-law prefactor to the
essential singularity \cite{Bray}, but this is not present in the
short-range case.  A disordered ferromagnet can be thought of as an
ensemble of weakly interacting, finite-sized ferromagnetic clusters.
When $T\sim T_G^-$, only large clusters contribute to the overall
magnetization. For each large cluster of linear size $L$, the smallest
Yang-Lee zero is of the order of $\theta\sim 1/mL^d$, where $m\sim
(T_G-T)^\beta$ is the magnetization per spin of the cluster and $d$ is
the spatial dimension of the system. The probability of a spin
belonging to a cluster of size $L$ follows the Poisson distribution,
\ie prob $\sim \exp(-cL^d)$. As a result, large clusters contribute to
$g(\theta,t)$ in the form of Eqn.~(\ref{Bray_form}), where $A(t)\sim
(T_G-T)^{-\beta}$ and $\beta$ is the usual exponent for pure
ferromagnets.

For $T<T_c$, the system is in its ferromagnetic phase with nonzero
magnetization, implying a nonzero value of $g(\theta=0,t)$\cite{Yang}.
This requires $A(t)=0$ at $T=T_c$ to counteract the essential
singularity at $\theta=0$. For $T\sim T_c^+$, we can expand $A(t)$ and
approximate it by $A(t)\sim t^{2-\beta_r}$.  This accounts for the
asymptotic behavior of $A(t)$.

With this form of $g(\theta,t)$, the scaling behavior of $M(h,t)$ can
be obtained by substituting Eqn.~(\ref{Bray_form}) into
Eqn.~(\ref{int_eqn}) and making a change of variables, $y=A(t)/\theta$,
yielding
\begin{equation}\label{no_power}
\frac{M(h,t)}{\mu} = \frac{g_0 A(t)}{2 h} \int_{A(t)/\pi}^{\infty} dy
\frac{\exp(-y)}{y^2+(A(t)/2h)^2},
\end{equation}
where we defined $h\equiv \mu H/k_BT_c$.  Eqn.~(\ref{no_power}) can be
written in terms of the exponential integral, $E_1(x)\equiv
\int_x^\infty e^{-t}/t \, dt$, as
\begin{equation}\label{m_as_E1}
\frac{M(h,t)}{\mu} = -g_0 \Im[\exp(i A(t)/2h) E_1(A(t)/\pi +
iA(t)/2h) ] .
\end{equation}
We expand
Eqn.~(\ref{m_as_E1}) about $A=0$, because $A(t) = A_0 t^{2-\beta_r}$ is
asymptotically small in the critical region, resulting in
\begin{equation}\label{true_m}
\frac{M(h,t)}{\mu} = -g_0 e^{-A(t)/\pi}
\times{\Im}[e^{iA(t)/2h}E_1(iA(t)/2h)].
\end{equation}
This implies an approximate scaling form
\begin{equation}\label{m_collapse}
\frac{M(h,t)}{\mu} = \exp(-A(t)/\pi)\left[ f_M(A(t)/h) + O(h)\right],
\end{equation}
where
\begin{equation}
f_M(x) = -g_0 {\Im}[\exp(ix/2)E_1(ix/2)],
\end{equation}
and the corrections of $O(h)$ involve exponential integrals of $A/h$.
This scaling prediction is valid in the limits $A\rightarrow 0$,
$h\rightarrow 0$, but the ratio $A/h$ has not been fixed by the
analysis so far.

\bigskip
\noindent {\it Analysis of magnetization data:-\/} We now analyze the
experimental data on
$\mathrm{La_{0.7}Ca_{0.3}MnO_3}$\cite{Salamon02,Salamon03}, using the
above results, to see if the data are consistent with the presence of a
Griffiths phase. Fig. (\ref{data_collapse}) shows the predicted data
collapse of the magnetization with fitted values $T_c=218K$,
$g_0=0.5\pm 0.05$, $A_0=0.5 \pm 0.05$ and $\beta_r=0.8\pm 0.05 $.  The
error bars were obtained by estimating the best fit visually.  The
figure also shows the agreement between the theoretically-predicted
universal scaling function and the collapsed data.  Overall, the data
scale quite well, and the scaling function of the collapsed data are
close to that of the theory, except near the turning point of the
curve, where corrections to the leading order ansatz we have used for
Eqn.~(\ref{m_collapse}) become important, and the data are no longer in
the asymptotic limits $A\rightarrow 0$ and $h\rightarrow 0$.  In order
to show this clearly, we calculate the asymptotics of the scaling
function, in the limits $A/h\rightarrow 0$ and $A/h\rightarrow \infty$,
where the leading terms in the magnetization can be calculated
systematically.

\begin{figure}[t]
\begin{center}
\includegraphics[width= .45\textwidth]{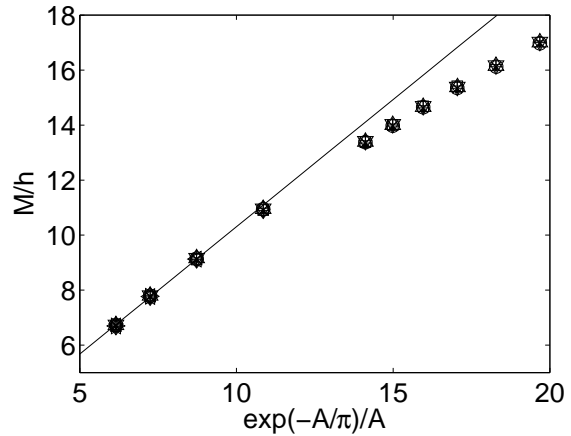}
\caption{Experimental verification of the asymptotic behavior of the
magnetization for $A/h\rightarrow \infty$.
The data show a linear dependence of $M/h$ on $\exp(-A/\pi)/A$ with
slope $0.93$ as predicted by Eqn.~(\ref{m_large_asymptotics}).  The
range of values of $A$ plotted is $0.05<A<0.155$. Legend: $\cdot$
$h=0.0006$, $\times$ $h=0.0011$, $\circ$ $h=0.0017$, $\bigtriangleup$
$h=0.0023$, $\bigtriangledown$ $h=0.0028$,  $+$ $h=0.0034$, $\ast$
$h=0.0045$, $\Box$ $h=0.0057$, $\diamond$ $h=0.0072$, $\triangleleft$
$h=0.0090$, $\triangleright$ $h=0.0122$, $\star$ $h=0.0141$ and the
solid line shows the linear prediction of Eqn.
~(\ref{m_large_asymptotics}).}
\label{large_a_over_h}
\end{center}
\end{figure}

\bigskip
\noindent {\it Asymptotic behavior:-\/} From Eqn.~(\ref{m_as_E1}), the
asymptotic behavior of $M(h,t)$, in the limit $A/h\rightarrow \infty$,
is given by
\begin{equation}
\frac{M(h,t)}{\mu g_0}\rightarrow \frac{2h}{A}e^{-A/\pi}
\left[1-\frac{4h^2}{\pi^2}-\frac{8h^2}{\pi A}+O\left(\frac{h^2}{A^2}\right)\right].
\label{m_large_asymptotics}
\end{equation}
showing that the small field susceptibility, $M/h$, in the limit
$A/h\rightarrow \infty$, depends linearly on $\exp(-A/\pi)/A$.
This prediction recovers Curie-Weiss-like behavior, as verified by the
experimental data shown in Fig.~(\ref{large_a_over_h}).
The experimental data shows a slope of $0.93$, consistent with the
fitted range of values for $0.9 < 2g_0 < 1.1$.

\begin{figure}[t]
\begin{center}
\includegraphics[width= .45\textwidth]{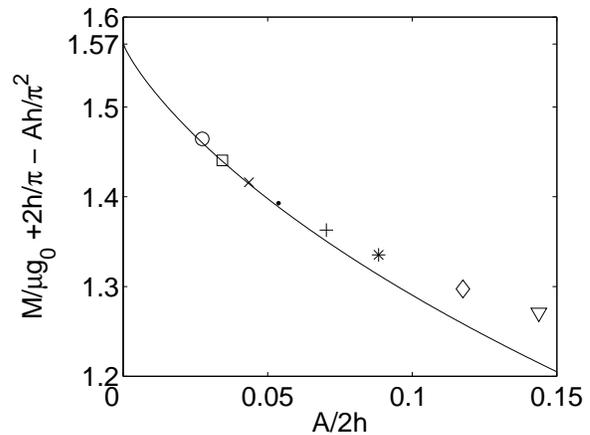}
\caption{Asymptotic behavior of the magnetization for $A/h\rightarrow
0$, showing that the data follow Eq. (\ref{m_small_asymptotics}), and
exhibit a discontinuity $M\rightarrow \mu g_0\pi/2$ in the limit
$h\rightarrow 0$.  The data shown here are within the limit $A/h<0.3$
and $h<0.0072$. Legend: $\bigtriangledown$ $h=0.0028$, $\diamond$
$h=0.0034$, $\ast$ $h=0.0045$, $+$ $h=0.0057$, $\cdot$ $h=0.0072$,
$\times$ $h=0.0090$, $\Box$ $h=0.0112$, $\circ$ $h=0.0141$.  Solid
line: theoretical prediction of Eq. (\ref{m_small_asymptotics}).}
\label{small_a_over_h}
\end{center}
\end{figure}

A profound difference between conventional and Griffiths ferromagnets
is found in the limit $A/h\rightarrow 0$.  It is well understood, for
conventional ferromagnets, that $M(h,t)\sim h^{1/\delta}\rightarrow 0$
as $h\rightarrow 0$; this is, however, not true in Griffiths
ferromagnets, as we can see by calculating the asymptotic behavior of
$M(h,t)$:
\begin{equation}\label{m_small_asymptotics}
\frac{M(h,t)}{\mu g_0}\sim  \frac{\pi}{2}+
\left(\ln\left(\frac{A}{2h}\right)+\gamma-1\right)\frac{A}{2h}
-\frac{2h}{\pi} + \frac{Ah}{\pi^2}
+O \left(\frac{A^2}{h^2}\right)
\end{equation}
as $A/h\rightarrow 0$, where $\gamma\sim 0.57722\dots$ is the
Euler-Mascheroni constant. This implies that $M(h,t)\rightarrow \mu
g_0\pi /2$ as $h$, $A/h\rightarrow 0$, {\it i.e.}, a discontinuity of
$M(h,t)$ at $h=0$ on the critical isotherm.  Experimental support for
this surprising result is present in Fig. 3 of Ref. (\cite{Salamon03}),
which documents the increase of the exponent $\delta$ fitted to data
assuming the conventional scaling applies.  As the transition
temperature is lowered by increasing disorder, the inferred value of
$\delta$ rises to as much as 16.9, inconsistent with any known
universality class, but indicative of a very rapid and dramatic rise in
magnetization.  In order to test the precise predictions made here, we
show in Fig. (\ref{small_a_over_h}) the experimental verification of
Eqn.~(\ref{m_small_asymptotics}), where the function $M/\mu
g_0+2h/\pi-Ah/\pi^2$ is plotted against $A/2h$.
The data points satisfying the criteria $h<0.0072$ and $A/h<0.3$ are shown in
Fig. (\ref{small_a_over_h}). All the parameters were determined
previously according to the data collapse in Fig. (\ref{data_collapse}),
so that we have not made any additional fitting. The experimental data
approach the theoretical curve as $A/h\rightarrow 0$, and moreover tend
to the universal number $\pi/2$ as dictated by
Eqn.~(\ref{m_small_asymptotics}).  We conclude that the data are
consistent with the prediction that $M(h,t)$ is discontinuous at $h=0$
in the limit $A/h \ll 1$, a prediction which follows from the essential
singularity characterizing the Griffiths transition\cite{Griffiths}.

\medskip
\noindent {\it Heat Capacity:-\/} We conclude with a brief discussion
of the heat capacity, $C(h,t)$. We can integrate
Eqn.~(\ref{m_collapse}) to obtain the free energy, $F(h,t)$ and thence
an expression for $C(h,t)$.  There is no prediction of data collapse,
due to the interference from the exponential terms,in agreement with
our failure to obtain data collapse from the data. This form contains
singular terms of the form
\begin{equation}
\label{Eqn:exponential_free_energy}
\int d\theta\,  e^{-A/\theta} \ln (4h^2 + \theta^2) \sim \exp (-A/h)
\end{equation}
where the integral is restricted to the neighbourhood of the origin
where Eqn. (\ref{Bray_form}) is valid and the upper limit is assumed to
scale with $h$, leading to the estimate of the essential singularity.
Similar terms have also been predicted in Ref. (\cite{Dotsenko06}), but
more than the leading term must be retained in order to consistently
compute the magnetization.

In conclusion, we have argued that the essential singularity of the
Griffiths phase leads to novel features in the critical behavior,
including a discontinuity of magnetization in external field.  These
features are reproduced to the accuracy expected from our lowest order
theoretical predictions by high quality experimental data from
$\mathrm{La_{0.7}Ca_{0.3}MnO_3}$, and lead to a consistent description
of its critical behavior, supporting the identification of this
material as a Griffiths ferromagnet.

\bibliographystyle{apsrev}

\bibliography{griffiths}

\end{document}